\documentclass[preprintnumbers, floatfix, preprintnumbers, letterpaper, superscriptaddress,nofootinbib, twocolumn]{revtex4}
\pdfoutput=1
\usepackage{graphicx}
\usepackage{microtype}
\usepackage{amsmath}
\usepackage{amssymb}
\usepackage{subfigure}
\usepackage{hyperref}
\usepackage{url}
\usepackage{xcolor}
\usepackage{color}
\usepackage{mathrsfs}
\usepackage{calrsfs}
\usepackage{amsfonts}
\usepackage{latexsym}
\usepackage{ragged2e}
\usepackage{epsfig}
\usepackage{textcomp}

\usepackage{caption}
\DeclareCaptionJustification{justified}{\leftskip=0pt \rightskip=0pt \parfillskip=0pt plus 1fil}
\definecolor{vividviolet}{rgb}{0.62, 0.0, 1.0}
\definecolor{amaranth}{rgb}{0.9, 0.17, 0.31}
\definecolor{palatinateblue}{rgb}{0.15, 0.23, 0.89}
\definecolor{brightpink}{rgb}{1.0, 0.0, 0.5}
\definecolor{cornflowerblue}{rgb}{0.39, 0.58, 0.93}
\definecolor{deepcarminepink}{rgb}{0.94, 0.19, 0.22}
\definecolor{radicalred}{rgb}{1.0, 0.21, 0.37}

\hypersetup{ linktoc=all,
    colorlinks, linkcolor={palatinateblue},
    citecolor={brightpink}, urlcolor={amaranth}
}

\graphicspath{{Images/}}

\makeatletter
\def\@fnsymbol#1{{\ifcase#1\or \textleaf  \else\@ctrerr\fi}}
	\makeatother

\begin{document}

\title{Generalized Uncertainty Principle, Black Holes, and White Dwarfs: \\A Tale of Two Infinities}

\author{Yen Chin \surname{Ong}}
\email{ycong@yzu.edu.cn}
\affiliation{Center for Gravitation and Cosmology, College of Physical Science and Technology, Yangzhou University, Yangzhou 225009, China}
\affiliation{School of Physics and Astronomy,
Shanghai Jiao Tong University, Shanghai 200240, China}

\begin{abstract}
It is often argued that quantum gravitational correction to the Heisenberg's uncertainty principle leads to, among other things, a black hole remnant with finite temperature. However, such a generalized uncertainty principle also seemingly removes the Chandrasekhar limit, i.e., it permits white dwarfs to be arbitrarily large, which is at odds with astrophysical observations. We show that this problem can be resolved if the parameter in the generalized uncertainty principle is negative. We also discuss the Planck scale physics of such a model.
\end{abstract}

\maketitle

\section{Introduction: Generalized Uncertainty Principle}

Quantum mechanics has two essential features that make it so different from classical physics -- quantum entanglement and the uncertainty principle. 
In the presence of a strong gravitational field, one usually expects some quantum corrections to gravitational physics. On the other hand, possible gravitational correction to quantum mechanics is (arguably) less well-understood. However, in order to obtain more hints about the full quantum gravity theory, it is important to understand how quantum mechanics and gravity affect each other.
It has been argued that quantum gravitational corrections could lead to various generalized versions of the uncertainty principle. Such modifications have been obtained via various general considerations of gravity and quantum mechanics \cite{9301067,9305163,9904025,9904026}, as well as from string theoretical considerations \cite{5,6,7,8,9}. 

The simplest version of the generalized uncertainty principle (GUP) can be obtained by considering quantum states in which $\langle \hat{p} \rangle = 0$, via the deformed commutation relation
\begin{equation}
[\hat{x},\hat{p}] = i\hbar \left[1+\alpha\left(\frac{\hat{p}}{2M_p}\right)^2\right].
\end{equation}
This leads to a GUP of the form
\begin{equation}\label{GUP}
\Delta x \Delta p \geqslant \frac{1}{2} \left[\hbar + \frac{\alpha L_p^2 \Delta p^2}{\hbar}\right],
\end{equation}
where $\alpha$ is the GUP parameter, $M_p$ the Planck mass, and $L_p$ the Planck length. For an attempt to further understand quantum mechanics with GUP, see \cite{1602.00608}. GUP implies the existence of a minimal length scale. See the comprehensive review by Hossenfelder \cite{1203.6191}. Note that string theory does not always lead to such a GUP \cite{yoneya}.

The GUP parameter is often taken to be of order unity in theoretical calculations, so that modification to the uncertainty principle only becomes obvious at the Planck scale. However, one could also treat Eq.(\ref{GUP}) phenomenologically, and seek experimental and observational bounds of the parameter $\alpha$. There have been many approaches to constrain $\alpha$ since the early work of Brau \cite{9905033}, which looked at the spectrum of hydrogen atom, but these bounds are rather loose. A short review of the various approaches can be found in the work of Scardigli and Casadio \cite{1407.0113}. Since then, a few other interesting works have tried to constrain $\alpha$ by using various means, from cold atoms \cite{1607.04353} to gravitational waves \cite{1610.08549v4} (see also \cite{1804.03620}), and atomic weak equivalence principle test \cite{1704.02037}. For example, $\alpha$ as large as $10^{34}$ is consistent with the Standard Model of particle physics up to 100 GeV \cite{1607.04353}. Tunneling current measurement yields a much lower bound: $\alpha \leqslant 10^{21}$ \cite{0810.5333}.

It has been argued that GUP leads to a generalized Hawking temperature, 
\begin{equation}
T=\frac{Mc^2}{4\alpha\pi}\left(1-\sqrt{1-\frac{\alpha \hbar c}{GM^2}}\right),
\end{equation}
which agrees very well with the usual expression for a \emph{large} asymptotically flat Schwarzschild black hole, $T=\hbar c^3/(8\pi GM)$, where Boltzmann constant $k_B=1$. However, as the black hole evaporates and becomes smaller in size, it eventually stops shrinking at around $M \sim M_p$, assuming that $\alpha \sim O(1)$. That is, GUP naturally leads to a black hole remnant \cite{pc}. Without GUP, Hawking temperature blows up, $T \to \infty$, as $M \to 0$, which is unsatisfactory. GUP ``cures'' this infinity. One may understand this in another way: ordinary black hole thermodynamics predicts temperature increases without bound as $M$ gets smaller. However, new physics might kick in once the energy is high enough, which prevents the black hole from evaporating further. The idea of black hole remnants dates back to Aharonov-Casher-Nussinov \cite{ACN1987}. See Chen-Ong-Yeom \cite{1412.8366} for a recent review.

What happens if we apply GUP to other physics, not just evaporating black holes? Since everything is quantum mechanical in nature, we can apply quantum physics to any system\footnote{ ``\emph{Once we have bitten the quantum apple, our loss of innocence is permanent.}'' -- R. Shankar \cite{shankar}} (of course, in the $\hbar \to 0$ limit this should reduce back to classical results). To have any hope of seeing the effects of GUP, one should choose a system in which gravity is sufficiently strong, thus a natural arena is compact star -- neutron star or white dwarf \cite{1301.6133, 1512.04337, 1512.06356}.  However, applying GUP to white dwarfs lead to a peculiar property: the Chandrasekhar limit no longer exists, thus arbitrarily large white dwarfs are seemingly allowed \cite{1512.06356}, but see caveats below. (See also \cite{1803.06640}, which discusses quantum gravity corrrections to white dwarf dynamics.) This is at odds with observations. Thus, while avoiding an infinity in black hole physics, GUP \emph{causes} another infinity in white dwarf physics.  In this work, we revisit this issue and find that choosing the GUP parameter to be negative resolves this problem elegantly, indeed \emph{both infinities are avoided by such a choice}. Our calculations are heuristic, focusing on the essence of the underlying physics -- the role of uncertainty principle in providing a degeneracy pressure to counteract gravitational collapse.

\section{Heisenberg's Uncertainty Principle and White Dwarfs}\label{I}

Let us first review the standard application of the usual Heisenberg's uncertainty principle to white dwarfs. For GUP the technique we shall employ is very similar, so it is good to be somewhat detailed in this review.

The Heisenberg's uncertainty principle $\Delta x\Delta p \gtrsim \hbar$ allows us to study some properties of white dwarfs, which are stellar remnants supported by electron degeneracy pressure. To this aim, let us consider the total kinetic energy of a non-relativistic white dwarf:
\begin{equation}\label{KE}
E_k = \frac{N\Delta p^2}{2m_e}\sim \frac{N\hbar^2}{(\Delta x)^2 2m_e},
\end{equation}
where $m_e$ is the electron mass. We have used the estimate $\Delta x \Delta p \sim \hbar$. Let the volume of the star be $V$ and $N$ the number of electrons (we assume the density to be homogeneous), and define number density $n:=N/V=M/(m_e V)$, where $M$ is the total mass of the star. We have $\Delta x \sim n^{-1/3}= (V/N)^{1/3}$ -- this is roughly the room an electron can move while being squeezed together. Their jittering motion gives rise to a huge $\Delta p$. With the relation $V \sim R^3$, where $R$ is the radius of the star, we now rewrite Eq.(\ref{KE}) as:
\begin{equation}
E_k = \frac{N\hbar^2n^{\frac{2}{3}}}{2 m_e} = \frac{N\hbar^2}{2 m_e} \left(M^{\frac{2}{3}}m_e^{-\frac{2}{3}}R^{-2}\right)=\frac{M^{\frac{5}{3}}\hbar^2}{2 m_e^{\frac{8}{3}} R^2}.
\end{equation}
To be at equilibrium, i.e., to withstand gravitational collapse, we must balance the kinetic energy with gravitational binding energy\footnote{For a spherical mass of uniform density, the gravitational binding energy is $E_g = -3GM^2/5R$.}:
\begin{equation}
|E_g|  \sim \frac{GM^2}{R} = E_k,
\end{equation}
which leads to 
\begin{equation}
R \sim \frac{\hbar^2}{2m_e^{\frac{8}{3}}G M^{\frac{1}{3}}}.
\end{equation}
Crucially, $R \propto M^{-\frac{1}{3}}$, or equivalently $M \propto 1/V$. This is the property of degenerate matter -- the more massive the star is, the smaller it gets. As the star gets smaller and smaller, eventually the movement of the electrons becomes relativistic. For relativistic regime, let us re-do the calculation, this time with relativistic kinetic energy
\begin{equation}
E_k = N(\gamma -1)m_ec^2, ~~ p=\gamma m_e v.
\end{equation}
With $v \sim c$, we have 
\begin{equation}
\Delta pc - m_e c^2 \sim \frac{\hbar c}{\Delta x} - m_e c^2 \sim \left(\frac{M}{m_e V}\right)^{\frac{1}{3}}\hbar c - m_e c^2.
\end{equation}
For $\Delta p$ large enough to withstand gravitational collapse, the rest mass term is negligible and one obtains
\begin{equation}
\frac{GM^2}{R} \sim N\left[\left(\frac{M}{m_e V}\right)^{\frac{1}{3}}\hbar c\right] \sim \frac{M^{\frac{4}{3}}\hbar c}{m_e^{\frac{4}{3}}R}.
\end{equation}
We note that $R$ gets canceled from both sides of the equation, and we are left with an expression for $M$:
\begin{equation}\label{Ch}
M_{\text{Ch}} \sim \frac{1}{m_e^2}\left(\frac{\hbar c}{G}\right)^{\frac{3}{2}}. 
\end{equation}
This is, modulo some constant overall factors obtained from more detailed and rigorous analysis\footnote{More precisely (see, e.g., Eq.(43) of \cite{chandra}), \begin{equation}M_\text{Ch}=\frac{\omega^0_3 \sqrt{3\pi}}{2}\left(\frac{\hbar c}{G}\right)^{\frac{3}{2}}\frac{1}{(\mu_e m_H)^2}, \notag\end{equation} where for a more realistic situation the star is not consisted purely of electrons, and $\mu_e$ denotes the average molecular weight per electron, while $m_H$ the mass of hydrogen atom. The constant coefficient $\omega^0_3 \approx 2.0182$ arises from the Lane-Emden equation.}, the famous \emph{Chandrasekhar limit}, also called the Chandrasekhar mass.

\begin{figure}[!h]
\centering
\includegraphics[width=3.0in]{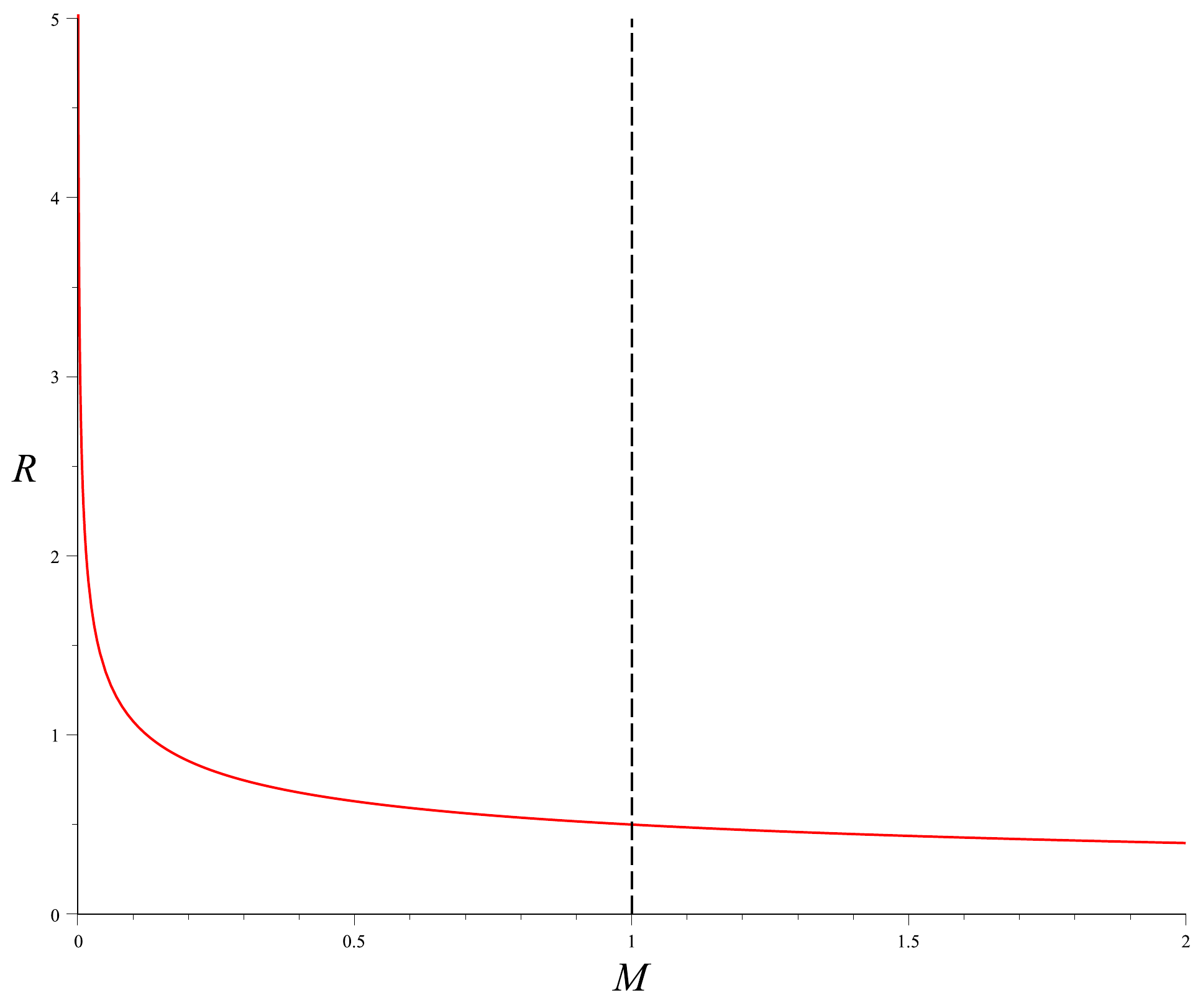}
\caption{The mass-radius relationship of a white dwarf: its size decreases as mass increases. The largest allowed mass for electron degeneracy pressure to balance gravitational collapse is the Chandrasekhar limit. Here we have set $G=c=\hbar=1$ and also take $m_e$ as unity (see sec.(\ref{III}) for an explanation). Thus the Chandrasekhar limit is $M_\text{Ch}=1$. The curve beyond $M > 1$ is therefore unphysical. \label{fig1}} 
\end{figure}

The physical interpretation is this: if we begin with a sufficiently large and non-relativistic white dwarf, as we add more and more mass\footnote{Astrophysically, this can be achieved in a close binary system, in which the white dwarf accrets hydrogen from its companion star, which often leads to flareup known as a ``nova''.},  the star becomes smaller as the electron degenerate pressure struggles to balance against the ever increasing gravity. The pressure of the white dwarf changes from $P \sim \rho^{\frac{5}{3}}$ to $P \sim \rho^{\frac{4}{3}}$ as it transits from non-relativistic regime into relativistic one, and the Chandrasekhar limit sets the bound beyond which no white dwarf can exist. See Fig.(\ref{fig1}). 

That is, adding more mass will no longer result in an even smaller star, instead gravitational collapse would occur (in realistic astrophysics, electrons will fuse with the remaining protons, creating a neutron star that counteracts gravitational collapse with neutron degeneracy pressure, but add enough mass and collapse into a black hole is inevitable).  Note that in a more rigorous analysis, one could make further distinction between relativistic and ultra-relativistic fermi gases. One obtains, for the relativitisc case, a curve that is strictly below the non-relativistic one, and terminates at the vertical line at $R=0$, so Chandrasekhar limit (which corresponds to \emph{ultra-relativistic} case) is not attainable. In our simple approach, we do not make such a distinction.

\section{How Generalized Uncertainty Principle Removes the Chandrasekhar Limit}\label{III}

With the generalized uncertainty principle (GUP), we have 
\begin{equation}
\Delta x \Delta p \sim \hbar + \alpha L_p^2 \Delta p^2/\hbar,
\end{equation}
which allows us to solve for $\Delta p$:
\begin{equation}\label{GUPp}
\Delta p \sim \frac{\hbar \Delta x}{2\alpha L_p^2} \left[1 \pm \sqrt{1-\frac{4\alpha L_p^2}{\Delta x^2}}\right].
\end{equation}
The sign in front of the square root can be chosen definitively (as will be done below), but let us keep both signs for now for clarity. 
Now we simply repeat the calculations in the previous section, which yields, for the non-relativistic case:
\begin{equation}
M^{\frac{5}{3}} \sim \frac{\hbar^2}{8 G m_e^{\frac{4}{3}} \alpha^2 L_p^4} R^3 \left(1 \pm \sqrt{1-\frac{4\alpha L_p^2M^{\frac{2}{3}}}{m_e^{\frac{2}{3}}R^2}}\right)^2.
\end{equation}
We are interested in the mathematical properties of the equation, so let us use the units in which $G=\hbar=c=1$, and furthermore take $m_e$ to be unity.
Note that by setting $G=\hbar=c=1$, we are using the Planck units in which $L_p=1=M_p^{-1}$. This means that the electron mass is actually a very small number ($\sim 4.1854 \times 10^{-23}$), \emph{not} unity. However, to keep the equation clean and not having to deal with very large (or very small) factors throughout, we have set $m_e=1$ purely for convenience. This does not affect the following analysis at the qualitative level -- ultimately for our heuristic approach, we are only interested in the behavior of the various mathematical functions, not its precise values. Physically, degenerate matter consisting of fermionic particles of some mass $m$ supports a compact star in the same manner no matter what $m$ is. So setting $m_e=1$ does not affect the underlying essential physics, and has the virtue of showing the results in a clean manner. We will restore $m_e$ for clarifty when needed.

Let us now consider the equation
\begin{equation}\label{MGUP}
M^{\frac{5}{3}} \sim \frac{R^3}{8\alpha^2} \left(1 \pm \sqrt{1-\frac{4\alpha M^{\frac{2}{3}}}{R^2}}\right)^2.
\end{equation}
If $\alpha$ is sufficiently small, we can expand to first order in $\alpha$ and obtain\footnote{If one fixes $\alpha$ instead, the expansion can be carried out by taking small $M$ limit. Keeping in mind that the aim is to recover the standard result without GUP: $R = 1/(2M^{\frac{1}{3}})$. In order for this to be true in the small $M$ limit, one must have $R \sim 1/M^{1/3} > M^{1/3}$, which justify the expansion.}:
\begin{equation}
M^{\frac{5}{3}}\sim \frac{R^3}{8\alpha^2}\left[1 \pm \left(1-\frac{2 \alpha M^{\frac{2}{3}}}{R^2}\right)\right]^2.
\end{equation}
Now we see that we should choose minus sign in front of the square root, so that this reduces to 
\begin{equation}
M^{\frac{5}{3}}\sim \frac{R^3}{8\alpha^2}\left[\frac{2\alpha M^{\frac{2}{3}}}{R^2}\right]^2 = \frac{M^{\frac{4}{3}}}{2R},
\end{equation}
that is, $R \sim M^{-\frac{1}{3}}$, as obtained using the usual Heisenberg's uncertainty principle.

\begin{figure}[!h]
\centering
\includegraphics[width=3.3in]{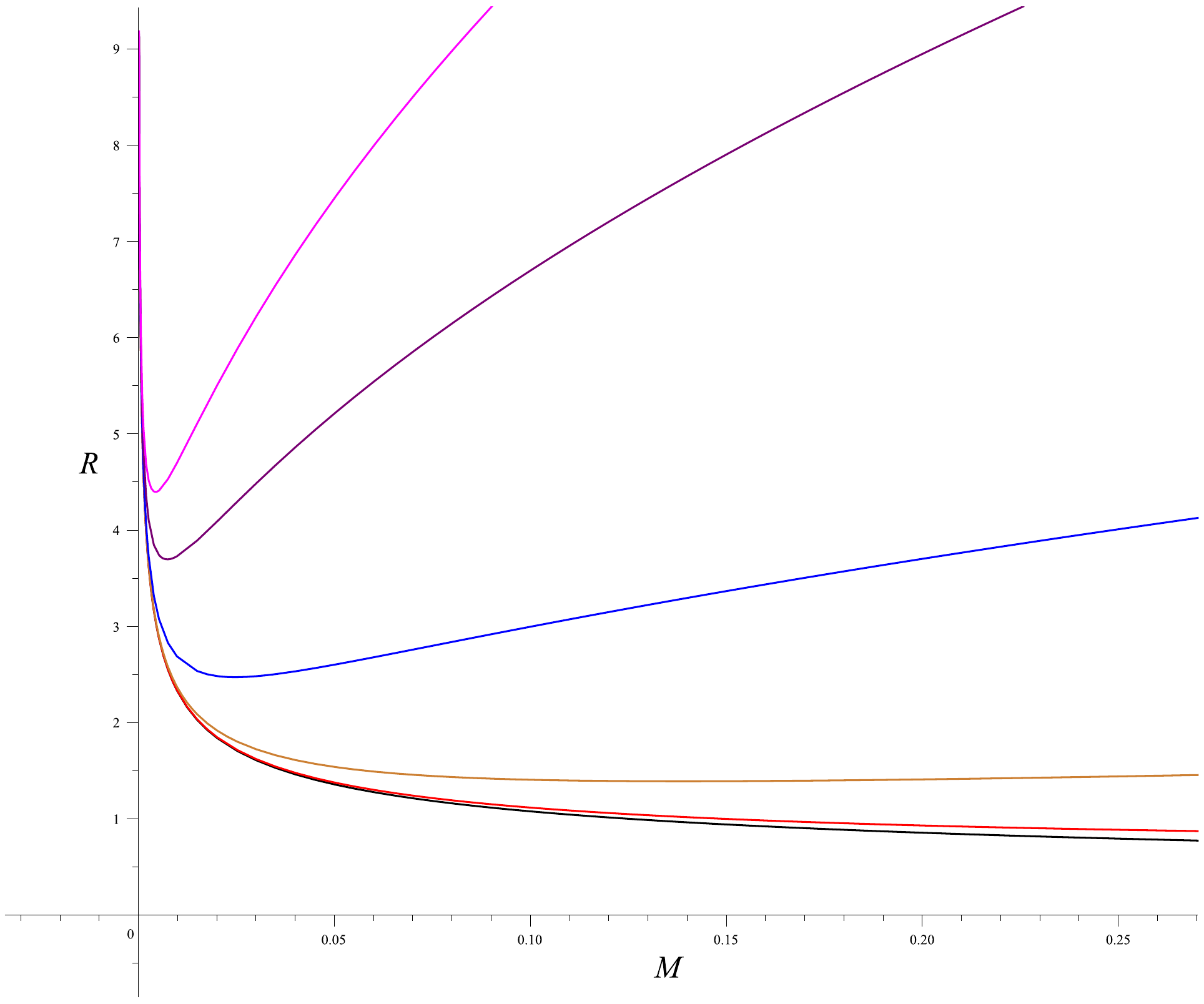}
\caption{The mass-radius relationship of a non-relativistic white dwarf with GUP correction: while its size initially decreases as mass increases, it eventually ``bounces'' and becomes unbounded in size. Here we have set $G=c=\hbar=1$ and also take $m_e$ as unity (see text for an explanation). The curves plotted, from top to bottom, correspond to $\alpha=100, 50, 10, 1, 0.1, 0$, respectively. The larger $\alpha$ is, the sooner the curve turns around. Crucially, no matter how small $\alpha$ is, the curve eventually turns around, unless $\alpha=0$.\label{fig2}} 
\end{figure}

Eq.(\ref{MGUP}), with the sign in front of the square root now fixed to be negative, can be solved analytically to obtain the stellar radius $R$ as a function of the mass $M$:
\begin{equation}
R(M):= \frac{1}{6}\left[\frac{1}{M^{\frac{1}{3}}} + \frac{1}{M}f(\alpha,M) + \frac{M^{\frac{1}{3}}(1+24\alpha M^{\frac{4}{3}})}{f(\alpha,M)}\right],
\end{equation}
where
\begin{flalign}\label{f}
f(\alpha,M):=&\left[M^2\left(216 \alpha^2 M^{\frac{8}{3}} + 24\sqrt{3} \alpha M^2 \sqrt{\alpha(27\alpha M^{\frac{4}{3}}+1)} \right.\right. \notag \\
&\left.\left.+36\alpha M^{\frac{4}{3}}+1\right)\right]^{\frac{1}{3}}.
\end{flalign}
We can check that as $\alpha \to 0$, we do obtain the correct limit $R \to 1/(2M^{\frac{1}{3}})$. 
We note that for large $R$, the leading term goes like $M^{\frac{5}{9}}$, that is, $\lim_{M\to \infty} R = \infty$. This is consistent with the conclusion in \cite{1512.06356}, which was obtained via a more rigorous analysis.
Fig.(\ref{fig2}) demonstrates this effect of the GUP parameter.

With hindsight, this is not surprising, as noted by Adler and Santiago \cite{9904026}, for $\alpha=1$, the GUP in Eq.(\ref{GUP}) is invariant under the ``inversion''
\begin{equation}
\frac{L_p \Delta_p}{\hbar} \leftrightarrow \frac{\hbar}{L_p\Delta_p }.
\end{equation}
This invariance implies that GUP relates high energies to low energies. That is to say,  the degenerate pressure produced by small values of $\Delta p$ should be the same as the one produced by large values of $\Delta p$. For $\alpha \neq 1$, the invariance is broken, but qualitatively the same effect holds.

\begin{figure}[!h]
\centering
\includegraphics[width=3.3in]{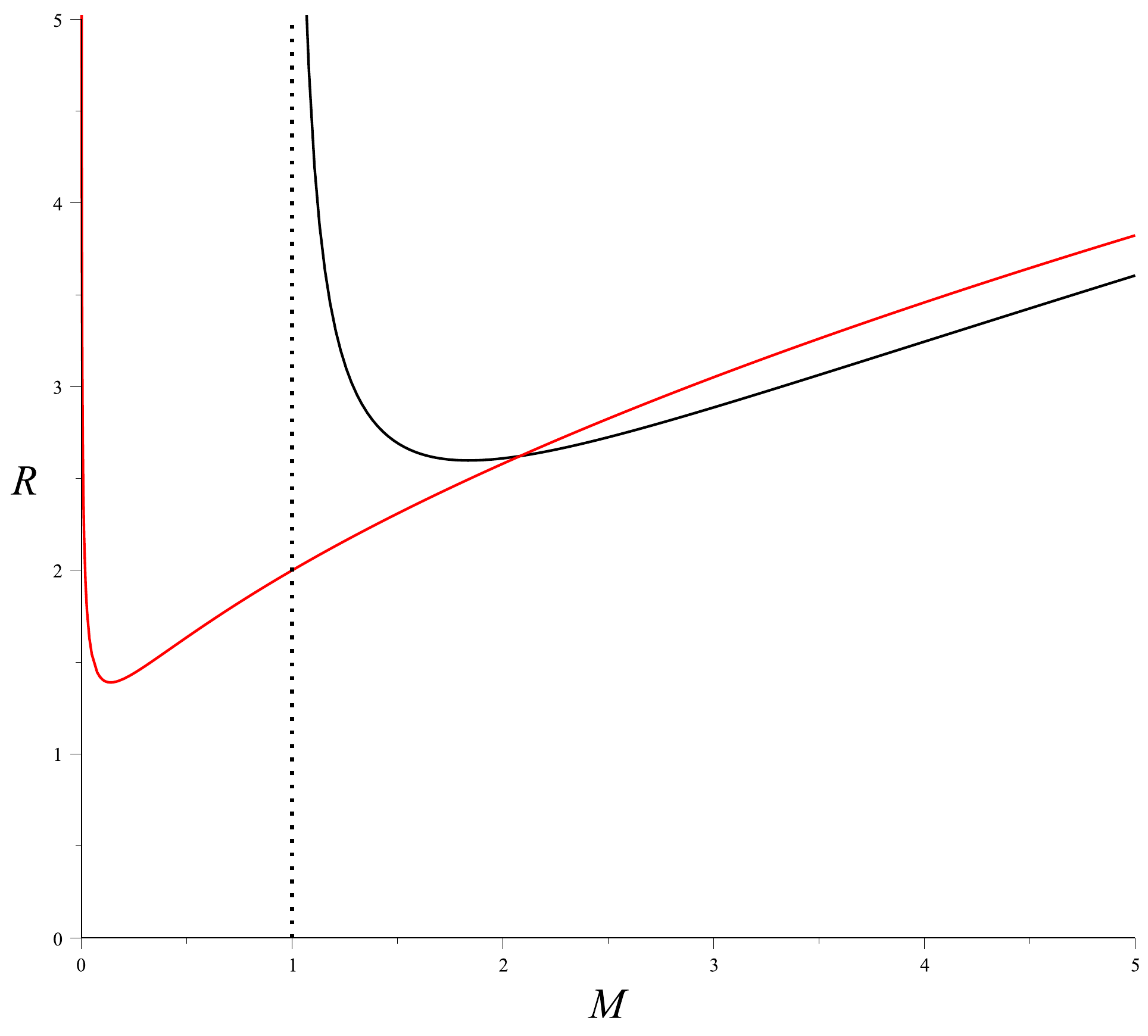}
\caption{The mass-radius relationship of a white dwarf with GUP correction $\alpha=1$. The red curve corresponds to non-relativistic case, while the black curve (which only exists for $M > 1$) corresponds to the relativistic case. Here we have set $G=c=\hbar=1$ and also take $m_e$ as unity (see text for an explanation). Both curves tend to infinity as $M$ does.   \label{fig3}}
\end{figure}

The relativistic case is very similar, but the expression for $R$ in terms of $M$ becomes considerably simpler:
The relation 
\begin{equation}\label{relGUP}
M^{\frac{4}{3}}=\frac{R^2}{2\alpha}\left(1-\sqrt{1-\frac{4\alpha M^{\frac{2}{3}}}{R^2}}\right)
\end{equation}
leads to
\begin{equation}\label{res1}
R(M) = \frac{M\sqrt{\alpha(M^{\frac{2}{3}}-1)}}{M^{\frac{2}{3}}-1}.
\end{equation}
It might be helpful at this stage to
restore $c, G, \hbar, m_e$. So this reads
\begin{flalign}
R(M) &= \frac{G M m_e\sqrt{\alpha G(GM^{\frac{2}{3}}m_e^{\frac{4}{3}}-c\hbar)}}{(GM^{\frac{2}{3}}m_e^{\frac{4}{3}}-c\hbar)c^2} \notag \\
&=\frac{\sqrt{\alpha}GM}{c^2}\frac{m_e^{\frac{1}{3}}}{\sqrt{M^{\frac{2}{3}}-M_\text{Ch}^{\frac{2}{3}}}}.
 \end{flalign}
Clearly this expression is not bounded above as $M \to \infty$. Consequently there is no Chandrasekhar limit. As mentioned in the introduction, this is problematic. Firstly, observations clearly show that the white dwarfs are bounded by the Chandrasekhar limit. In contrast, both the non-relativistic and relativistic curves of $R(M)$ are unbounded above, as illustrated by Fig.(\ref{fig3}). In addition, one notes that the domain of the Chandrasekhar limit curve is bounded below by $M_\text{Ch}$, given by Eq.(\ref{Ch}), which if we set $c,G,\hbar,m_e$ to unity, is equal to 1. This is because the square root term in Eq.(\ref{res1}) has to be positive, so $M > M_\text{Ch}$, and indeed Eq(\ref{res1}) diverges in the limit $M \to M_\text{Ch}^+$. This parallels the behavior of the non-relativistic GUP curve in that both curves tend to infinity as they approach the respective lower bounds ($M \to 0$ for the non-relativistic GUP curve). What this means is that for $M$ sufficiently close to $M_\text{Ch}$, the relativistic curve recovers the original Chandrasekhar limit. In other words, without GUP correction, the most extreme white dwarf allowed is that of $M_\text{Ch}$. However, with $\alpha > 0$, GUP correction allows arbitrarily large white dwarf, since the relativistic curve grows without bound as $M$ increases. 

It is rather perplexing as to how a small deviation -- no matter how small $\alpha$ is -- from the uncertainty principle causes such a large deviation for massive stars. This goes against our usual expectation that GUP correction should only be obvious when we reached the Planck scale, that is to say, white dwarfs should not be affected that much (if at all) by GUP correction.

Observationally, there have been a few candidates of  ``super-Chandrasekhar'' white dwarfs \cite{0609616, 1106.3510}, which seem to exceed the Chandrasekhar limit, but their masses are still of order $O(M_\text{Ch})$. The fact that GUP allows not just large white dwarfs, but \emph{arbitrarily} large ones, is disconcerting. Secondly, if we think in terms of gravitational collapse by adding more mass, what the result suggests is this: while without GUP correction, a white dwarf continues to get smaller in size until it hits Chandrasekhar limit beyond which gravitational collapse will again proceed, in the GUP corrected case, at some point the white dwarf resists collapse and actually \emph{bounces} back and continue to grow indefinitely. This is consistent with the discussion in \cite{1512.04337} around their equation for $R(M)$, i.e. Eq.(55) in their work, which is essentially the same as our Eq.(\ref{res1}). This means that black holes might not form due to GUP-enhanced degenerative pressure resisting the collapse. If true, this would go against observations which show a plethora of black holes in the Universe. 

Here, let us mention a caveat: in Fig.(\ref{fig3}), one might be tempted to consider also the limit imposed by the Schwarzschild radius, i.e., if $R < 2M$ one might expect the star to have collapsed into a black hole, and so whatever under the line $R=2M$ is irrelevant. However, in order to discuss this consistently, general relativistic correction has to be taken into account, and the (GUP-corrected) Tolman-Oppenheimer-Volkoff  equation would be required to fully model the stars. This is beyond the scope of the current work, but clearly deserved to be studied in more quantitative details. Our proposal below has the advantage in the sense that it restores the Chandrasekhar limit directly, without relying on black hole formation to effectively ``shield'' the novel GUP effect. In addition, in view of how surprising GUP effect can be, one cannot be entirely confident that the criteria for black hole formation (i.e. the Schwarzschild limit) is not affected as well.

It is possible that various other factors -- such as coulomb correction, angular momentum correction (compact stars tend to spin very fast), magnetic field (compact stars tend to have a large magnetic field) correction, and lattice energy --  may come into play to prevent white dwarfs from growing too massive \cite{1512.04337}. Nevertheless, it would be more convincing if one could resolve this problem entirely within the context of GUP physics. 
It is of course also possible that this tension with observations hints at GUP being incorrect (or not applicable in some regimes). However, due to its various virtues, we hope to still have some form of GUP, but without the aforementioned problems.  We propose a simple resolution: \emph{$\alpha$ should be taken to be negative}.

\section{How A Negative Sign Removes Both Infinities}

\begin{figure}[!h]
\centering
\includegraphics[width=3.3in]{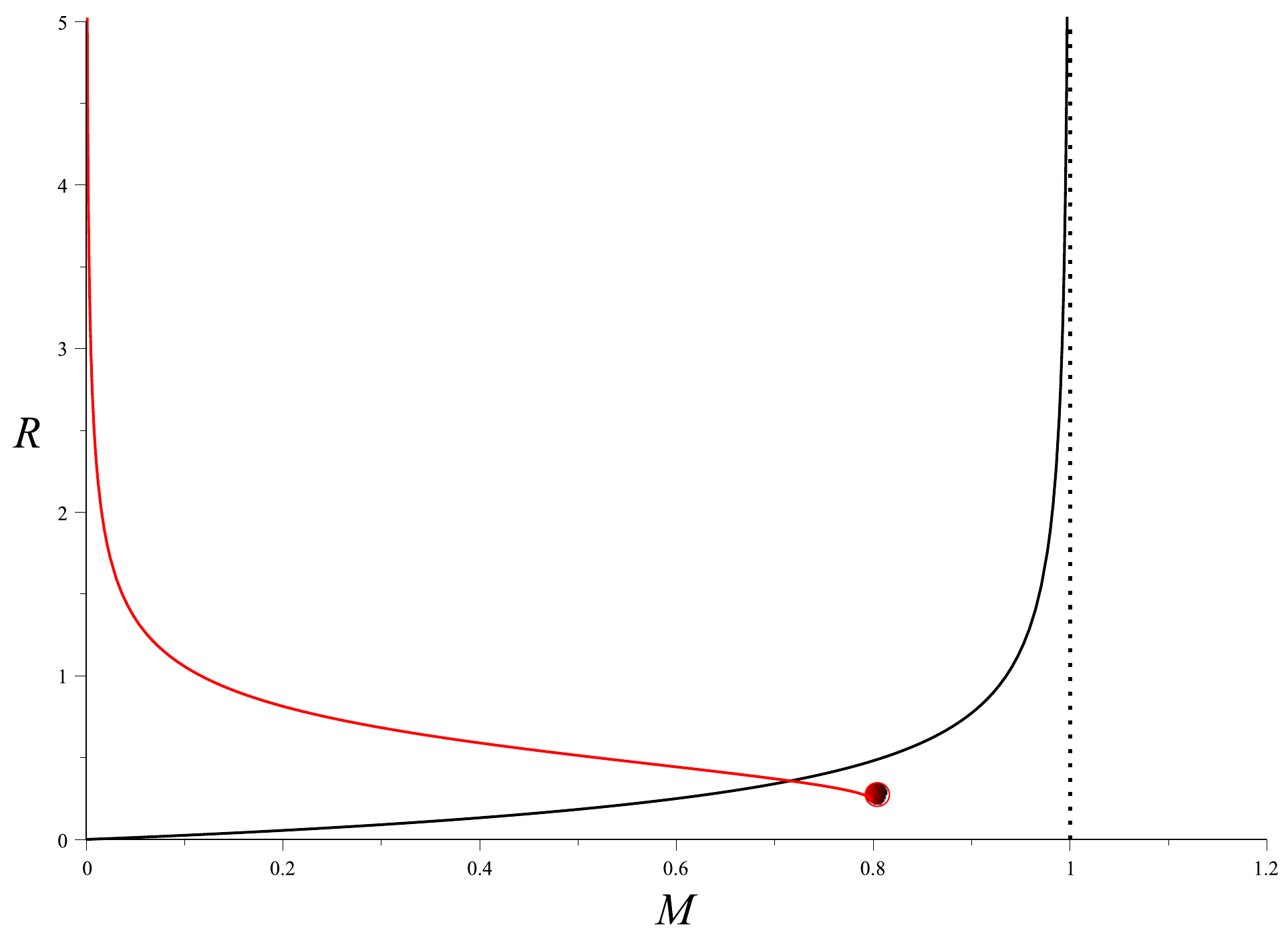}
\caption{The mass-radius relationship of a white dwarf with GUP correction $\alpha=-0.05$. The red decreasing curve corresponds to non-relativistic case, while the black increasing curve corresponds to the relativistic case, which asymptotes to $M=1$. The non-relativistic curve terminates at around $M=0.797$. Here we have set $G=c=\hbar=1$ and also set $m_e$ as unity (see text for an explanation).  \label{fig5}}
\end{figure}

\begin{figure}[!h]
\centering
\includegraphics[width=3.3in]{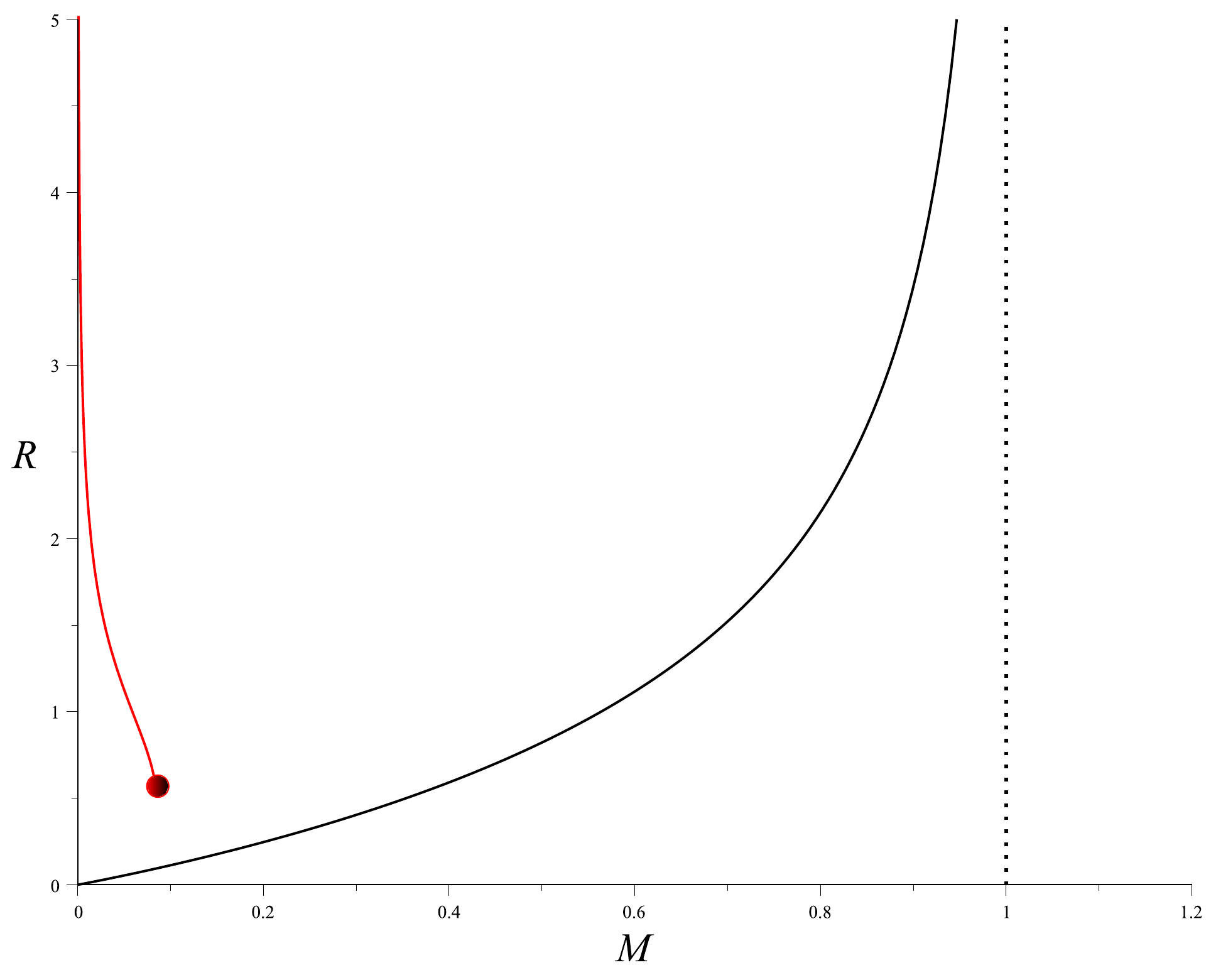}
\caption{The mass-radius relationship of a white dwarf with GUP correction $\alpha=-0.1$. The red decreasing curve corresponds to non-relativistic case, while the black increasing curve corresponds to the relativistic case, which asymptotes to $M=1$. The non-relativistic curve terminates at around $M=0.084$. Here we have set $G=c=\hbar=1$ and also take $m_e$ as unity (see text for an explanation).  \label{fig4}}
\end{figure}

As in turns out, if we consider $\alpha < 0$, then we will not have arbitrarily large white dwarfs. For example, setting $\alpha=-0.05$, we obtain the plots in Fig.(\ref{fig5}). Note that for the relativistic case, one now has to take the other root of Eq.(\ref{relGUP}):
\begin{equation}
R(M) = -\frac{M\sqrt{\alpha(M^{\frac{2}{3}}-1)}}{M^{\frac{2}{3}}-1},
\end{equation}
so that $R > 0$.
Again, it might be helpful at this stage to
restore $c, G, \hbar, m_e$, which yields
\begin{flalign}\label{restore}
R(M) 
&= -\frac{G M m_e\sqrt{-|\alpha| G(GM^{\frac{2}{3}}m_e^{\frac{4}{3}}-c\hbar)}}{(GM^{\frac{2}{3}}m_e^{\frac{4}{3}}-c\hbar)c^2} \notag \\
& = \frac{\sqrt{|\alpha|}GM}{c^2}\frac{m_e^{\frac{1}{3}}}{\sqrt{M_\text{ch}^{\frac{2}{3}}-M^{\frac{2}{3}}}}.
\end{flalign}
The plots obtained is then only a minor modification from the ones obtained via the usual Heisenberg's uncertainty principle, as one should expect from a \emph{small} GUP correction.
GUP has two effects: for the non-relativistic case, it terminates the curve at some $M_{\text{max}}$, beyond which no white dwarfs are allowed. Thus, with GUP one obtains a Chandrasekhar limit even when the star is not relativistic. Furthermore, for the relativistic case, instead of obtaining a vertical line at $M=1$, one now obtains a curve that asymptotes to $M=1$. 
This is really the Chandrasekhar mass given in Eq.(\ref{Ch}), as can be seen from Eq.(\ref{restore}). 
The vertical line $M=1$ can never be reached and serves as the relativistic Chandrasekhar mass. The existence of such a bound is due to the function $f(\alpha, M)$, defined in Eq.(\ref{f}), which becomes complex for sufficiently large $M$.
GUP bends the curve away from the line $M=1$ for small values of $R$. Increasing the value of $\alpha$ has the effect of terminating the non-relativistic curve at smaller value of $M$. For example, for $\alpha=-1$, as shown in Fig.(\ref{fig4}), the non-relativistic curve terminates before it cross the relativistic one. However the general trend is the same: \emph{white dwarfs are bounded above in size}. 

There is another piece of evidence that supports the choice of $\alpha < 0$. As mentioned by Moussa in \cite{1512.04337}, the current observation indicates that some white dwarfs
have smaller radii than theoretical predictions \cite{0604366,0610073,ironbox}. Moussa commented that this is not consistent with GUP (with $\alpha > 0$). We can see this in Fig.(\ref{fig2}), since all the curves are \emph{above} the curve obtained from the unmodified uncertainty principle, $R(M,\alpha=0)$. However, if $\alpha < 0$, then the curves \emph{are} lower than  $R(M,\alpha=0)$.

One might worry that choosing a negative GUP parameter will affect its application to black hole evaporation. Recall that the generalized Hawking temperature is 
$T = {pc}/(4\pi)$, where $p$ is taken to be the value of the RHS of  Eq.(\ref{GUPp}) by replacing $\Delta x = 2GM/c^2$, the Schwarzschild radius (assuming this is unchanged under GUP correction). When $\alpha > 0$ this automatically imposes an upper bound of temperature since temperature needs to be a real number: Hawking evaporation stops when $\Delta x = 2\sqrt{\alpha}L_p$, that is, as $M=\sqrt{\alpha}M_p$. If $\alpha \sim O(1)$, then the black hole stops evaporating when its size is of the order of Planck mass (though oddly in this simple model it continues to emit radiation at finite temperature).

Now, if we consider $\alpha < 0$, the square root does not impose a bound on the black hole size. That is, \emph{we no longer have a remnant}. However, the expression for the generalized Hawking temperature is still real and positive. For simplicity, we again set $G=c=\hbar=1$, and compare the two generalized Hawking temperatures:
\begin{equation}
T[\alpha>0] = \frac{M}{4\alpha\pi}\left(1-\sqrt{1-\frac{\alpha}{M^2}}\right), ~~\text{and}
\end{equation} 
\begin{equation}
T[\alpha<0] = -\frac{M}{4|\alpha|\pi}\left(1-\sqrt{1+\frac{|\alpha|}{M^2}}\right),
\end{equation} 
as well as the unmodified Hawking temperature, 
\begin{equation}
T[\alpha=0]=\frac{1}{8\pi M}.
\end{equation}
For large $M$, both $T[\alpha>0]$ and $T[\alpha<0]$ agree with $T[\alpha=0]$. However,
We see that for $M$ small enough, the dominant term under the square root of $T[\alpha <0]$ is $\alpha/M^2 \gg 1$, so that 
\begin{equation}\label{Talphan}
T[\alpha<0] \sim \frac{M}{4|\alpha|\pi} \frac{\sqrt{|\alpha|}}{M}=\frac{1}{4\pi\sqrt{|\alpha|}} < \infty.
\end{equation} 

\begin{figure}[!h]
\centering
\includegraphics[width=3.3in]{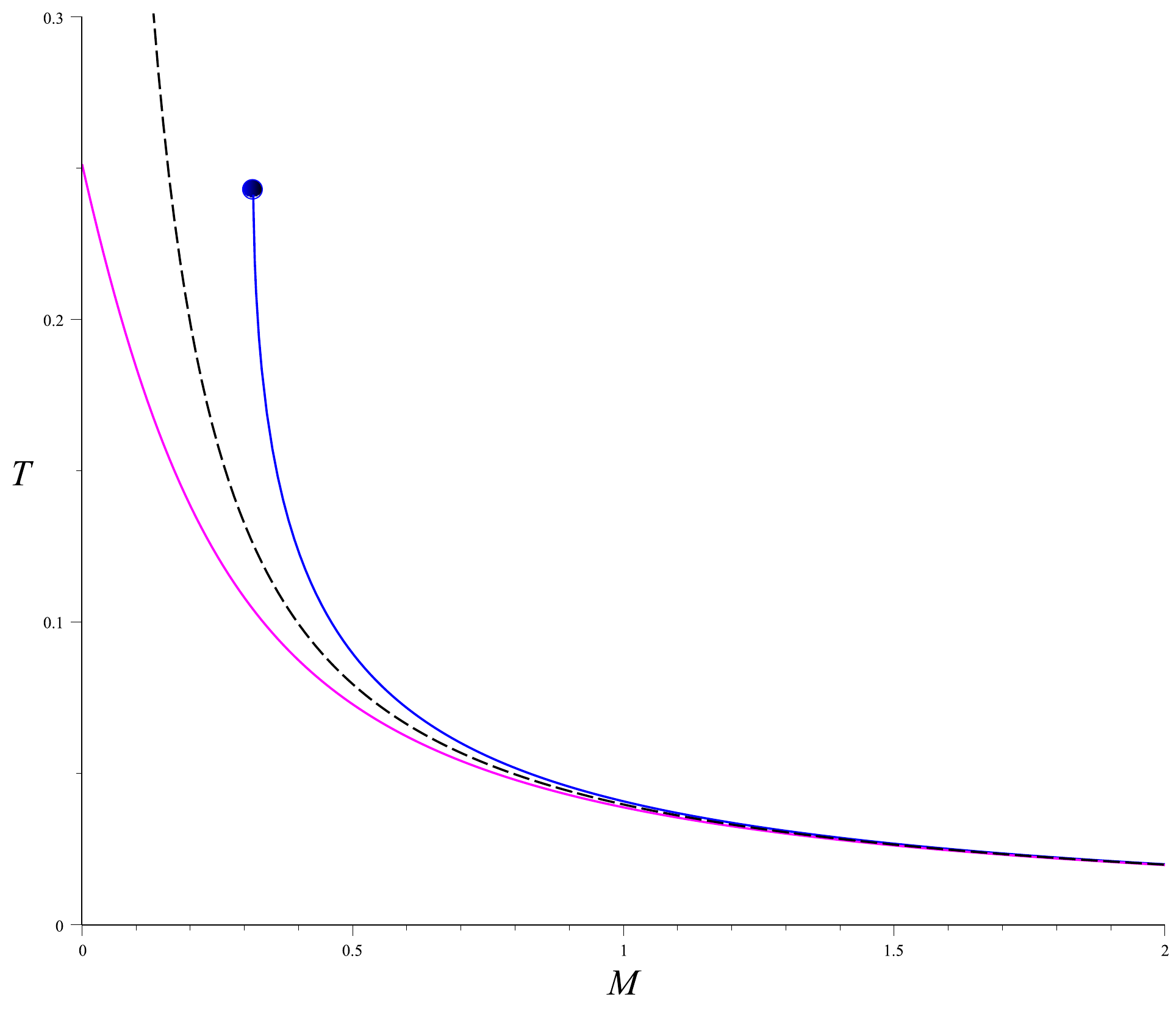}
\caption{The Hawking temperature of an asymptotically flat Schwarzschild black hole: without GUP correction, we obtained the middle curve, which increases without bound as $M \to 0$. The infinity is ``cured'' with GUP correction: in the usual consideration $\alpha > 0$,  the temperature curve terminates at around $M \sim \sqrt{\alpha}M_p$. This is shown by the right-most curve. If $\alpha < 0$, the GUP correction no longer imposes a lower bound on the black hole size. However, as the left-most curve shows, its temperature remains finite as the black hole shrinks down to zero size, without leaving behind any remnant.
\label{fig6}}
\end{figure}

Despite not having a lower bound for the black hole size, the Hawking temperature remains finite as the black hole evaporates down to zero size, as shown in Fig.(\ref{fig6}). In other words, choosing $\alpha <0$ \emph{also} prevents the temperature from blowing up, but it has the additional advantage over choosing $\alpha > 0$ since it also prevents white dwarfs from becoming too large.  

This of course does not mean that black hole remnants do not exist, but rather from GUP argument alone, if $\alpha < 0$, one does not obtain a remnant. After all, such simple model cannot be expected to capture all properties of the full quantum gravity theory. On the other hand, allowing black holes to shrink to zero size (and thus no more radiation thereafter) seems more satisfactory than having a remnant that,  while having a \emph{fixed} mass, continues to radiate at some finite temperature. We also note from Eq.(\ref{Talphan}) that we have a \emph{universal} bound of the Hawking temperature, which is independent of the black hole mass. That is, no matter the initial mass of the black hole, it always radiate down to a final temperature of $1/(4\pi \sqrt{|\alpha|})$ before completely evaporates away. 

It should be emphasized that negative values of $\alpha$ has been previously discussed in the literature, though it is not as widely considered as the $\alpha >0$ case. In particular $\alpha < 0$ (and the lack of nonzero mass remnant) is consistent with the findings of Jizba-Kleinert-Scardigli \cite{0912.2253}, who derived a similar GUP assuming that the universe has an underlying crystal lattice-like structure. In addition, Scardigli and Casadio \cite{1407.0113} also showed that if one takes the generalized Hawking temperature, and make the reasonable assumption that one should be able to obtain it from Wick-rotating a deformed static Schwarzschild metric 
with metric coefficient
\begin{equation}
g_{tt}=-\left(1-\frac{2M}{r} + \varepsilon\frac{M^2}{r^2}\right),
\end{equation} 
then for $|\varepsilon| \ll 1$, we have 
\begin{equation}
\alpha = -4\pi^2\varepsilon^2\left(\frac{M}{2 M_p}\right)^2 < 0.
\end{equation}
If we take observations of white dwarfs seriously, then the fact that they are all around the Chandrasekhar limit (if not all bounded by it) \emph{by itself} already suggests that $\alpha$ should be negative, independent of the previous mentioned arguments.

\section{Discussion: The Sign of GUP Parameter}

In this work we have confirmed the previous results in the literature that GUP with positive parameter $\alpha$ allows arbitrarily large white dwarfs to exist. 
Although our analysis is less rigorous and largely heuristic, it provides a clean view of how GUP enters the physics of white dwarfs, by keeping the mathematics to the bare minimum.
In addition to arbitrarily large white dwarfs, we note that having the curve $R(M)$ turns around as $M$ is increased means that white dwarfs ``bounce'', which
suggests that gravitational collapse into black hole might not occur, though this requires more detailed analysis. Both of these are at odds with astrophysical observations. If anything, observational data suggests that some white dwarfs are smaller than theoretical expectation, which also contradicts the results found from applying GUP to white dwarfs. \emph{We resolve all these problems with a simple proposal: the GUP parameter should be negative}. Such a choice simultaneously maintains a finite temperature at the end of Hawking evaporation, a virtue that GUP with $\alpha > 0$ also enjoys (but leads to a black hole remnant). White dwarf physics thus provides an additional piece of evidence that supports the choice $\alpha < 0$, in addition to some other arguments previously found in the literature. 

The next question one should consider is: \emph{why} is $\alpha < 0$? Phenomenologically the usual choice is to set $\alpha > 0$, since in various ``derivations'' and thought experiments of GUP it seems more reasonable to have $\alpha > 0$. In some string theoretic considerations, for example,  $\alpha$ is the Regge slope parameter $\alpha' > 0$, which is related to the string length $\lambda_s$ by $\lambda_s^2 = \hbar a'$. It is therefore surprising if $\alpha$ does turn out to be negative; one would then need to understand the underlying reason. (We should emphasize that, as pointed out in \cite{0912.2253}, stringy GUPs are not rigorously derived, but instead deduced from high-energy
thought experiments involving string scatterings.) 

Let us also remark that if $\alpha < 0$, then the uncertainty is \emph{suppressed}. In fact if $|\alpha|$ is around unity, then as one approaches Planck scale, there exists a maximum $\Delta p$ such that $\Delta x\Delta p \geqslant 0$. It thus appears that physics becomes classical again at Planck scale, instead of some form of highly fluctuating spacetime foam as one usually expects to find. This phenomenon has been previously pointed out and discussed in \cite{0912.2253} and \cite{1504.07637}. Let us expound it further in this work: In fact, such a possibility of a ``classical'' Planckian regime was already explore in the literature, e.g., by considering $\hbar$ as a dynamical field that goes to zero in the Planckian limit \cite{1208.5874, 1212.0454}. Since in 4-dimensions, the gravitational constant is $G=\hbar c/M_p^2$, if we fixed Planck mass, then $\hbar \to 0$ is equivalent to $G \to 0$. Therefore, ``classicalization'' happens also in the context of, e.g., asymptotic safe gravity \cite{0610018}, and other scenarios in which gravity is weakened at high energy, e.g. in the context of $f(R)$ gravity \cite{0911.0693}. In other words, the fact that  $\Delta x\Delta p \geqslant 0$ for $\alpha < 0$ is not all that peculiar\footnote{A suggestive piece of evidence is the singularity of a dilaton charged black hole,  see p.12 of \cite{9210119} by Horowitz (emphasis added):
``[...] the string coupling is becoming very weak near the singularity. As we have discussed, we have no right to trust this solution near the singularity, but its difficult to resist speculating about what it might mean if the exact classical solution had a similar behavior. \emph{It would suggest that, contrary to the usual picture of large quantum fluctuations and spacetime foam near the singularity, quantum effects might actually be suppressed.} The singularity would behave classically.''} -- it just shows that GUP can also serve as a simple phenomenological model for this kind of quantum gravity. \emph{Given our ignorance about Planckian physics, it is a strength of GUP that it could accommodate various scenarios with different signs of the GUP parameter}.

There are of course other possibilities. Firstly, it could be that $\alpha$ is indeed positive, but white dwarfs involve so much other physics that GUP effect is suppressed. Secondly, $\alpha > 0$ might not be a fixed number. In \cite{1408.3763}, it is suggested that for black hole complementarity principle to always hold, $\alpha$ should depend on the number of the underlying quantum fields. Thus, it is possible that $\alpha$ is a \emph{function}, which gives rise to sensible white dwarf physics. However this possibility seems unlikely, since for this to work $\alpha$ would need to be a function of the total mass $M$. 

In order to further understand how quantum mechanics and gravity influence each other, the generalized uncertainty principle should be further investigated to determine the sign of the GUP correction. We should also further investigate how a negative GUP parameter might affect other areas of physics beyond white dwarfs and black holes, say, perhaps in cosmology \cite{fawad}. At this point we should probably keep the options open: is it also possible that there is no universal sign for GUP correction, and whether $\alpha$ is positive or negative depends on some external factors of the physical system under consideration? Given that GUP physics is largely heuristic even as phenomenological models, one should certainly keep an open mind and explore various possibilities of its ``parameter space'' in search of viable choices that give sensible physics, such as those that do not yield arbitrarily large white dwarfs.

\begin{acknowledgments}
YCO thanks the National Natural Science Foundation of China (grant No.11705162) and the Natural Science Foundation of Jiangsu Province (No.BK20170479) for funding support. 
He also thanks Brett McInnes and Fawad Hassan for comments and suggestions. 
\end{acknowledgments}

\end{document}